\documentclass[preprint,showpacs,preprintnumbers,amsmath,amssymb,11pt,]{revtex4}

\usepackage{graphicx}
\usepackage{dcolumn}
\usepackage{bm}
\usepackage{ulem}

\begin{document}


\title{Raman Model Predicting Hardness of Covalent Crystals}

\author{Xiang-Feng Zhou,$^{1,3}$ Quang-Rui Qian,$^{3}$ Jian
Sun,$^{3,\dagger}$\footnotetext{$^\dagger$Present address: Lehrstuhl
f\"{u}r Theoretische Chemie, Ruhr-Universit\"{a}t Bochum, 44780
Bochum, Germany} Yongjun Tian,$^{2}$ and Hui-Tian Wang$^{1,3}$}
 \email{htwang@nju.edu.cn;htwang@nankai.edu.cn}
\affiliation{$^{1}$School of Physics, Nankai University, Tianjin 300071, China
\\$^{2}$ State Key Laboratory of Metastable Materials Science
and Technology, Yanshan University, Qinhuangdao 066004, China\\$^{3}$Nanjing National Laboratory of Microstructures
and Department of Physics, Nanjing University, Nanjing 210093,
China}

\date{\today}

\begin{abstract}
\noindent Based on the fact that both hardness and vibrational Raman
spectrum depend on the intrinsic property of chemical bonds, we
propose a new theoretical model for predicting hardness of a
covalent crystal. The quantitative relationship between hardness and
vibrational Raman frequencies deduced from the typical zincblende
covalent crystals is validated to be also applicable for the complex
multicomponent crystals. This model enables us to nondestructively
and indirectly characterize the hardness of novel superhard
materials synthesized under ultra-high pressure condition with the
\textit{in situ} Raman spectrum measurement.

\end{abstract}


\maketitle


Design and synthesis of new superhard materials are of great
interest due to their numerous applications.
\cite{R01,R02,R03,R04,R05} Hardness, as an important macroscopic
physical property, is understood as the resistance offered by a
given material to applied mechanical action. For a crystalline
material, hardness is an intrinsic property. Its prediction, from
the microscopic electronic structure, is a crucial issue and a
powerful challenge in condensed matter physics and materials
science. Recently, the microscopic model connecting hardness with
the nature of chemical bond has shed light on the quantitative
estimations of other macroscopic properties.\cite{R06,R07,R08} To
predict hardness using the above models, the exact crystal structure
must be known. Experimentally, the ultra-high-pressure technique
such as diamond anvil cell (DAC) is extensively used to explore new
superhard materials.\cite{R09,R10,R11,R12} Due to the limited size
of the sample synthesized in DAC, it is often very difficult to
either conclusively determine the exact atomic arrangement or the
hardness for some newly synthesized samples. These quantitative
models are thereby limited on some practical applications. However,
the \textit{in situ} measurements of Raman and infrared equipped on
DAC can provide the diagnostic electronic features of the phase.

Raman spectroscopy is commonly used in chemistry, since vibrational
information is specific for the chemical bonds in a crystal. It
therefore provides a fingerprint for identifying the crystal. The
Raman scattering process involves the change in susceptibility while
the infrared absorption connects with the change in dipole moment.
The Raman modes include vibrational, rotational, and librational
modes, while the infrared absorption mode involves translational
mode only. Furthermore, the vibrational Raman modes can also be
classified into the longitudinally optical (LO) and transversely
optical (TO) modes, originating from the tensing/compressing and
bending of chemical bonds, respectively.\cite{R05} Based on the
microscopic understanding, hardness is naturally the resistance of
chemical bond per unit area to indenter.\cite{R06} In the
experimental measurement by using the indenter, the hardness value
is determined from the ratio of the load to the indentation
area.\cite{R05} This resistance is related to the
tensing/compressing and bending, but not the rotating, librating or
translating of chemical bonds because the bond length will keep the
same for the translation and rotation (libration) modes, it wouldn't
embody the effective bond strength if considering simulation of the
indentation, the contacted bonds would be deformed or broken.

In the previous works, the resistant force of bond can be
characterized by energy gap,\cite{R06} reference energy \cite{R07}
and bond electronegativity.\cite{R08} In our understanding, the
\textit{vibrational} Raman frequency/energy can also be used to
describe the resistant force, due to the reasons as mentioned above.
In this work, we create a quantitative relationship between hardness
and \textit{vibrational} Raman modes in nonresonant first-order
Stokes Raman spectrum. Raman spectra find other important
application on the prediction of hardness.

Inasmuch as the \textit{vibrational} Raman frequency of a chemical
bond increases as the force constant of a chemical bond increases,
the \textit{vibrational} Raman frequency embodies is in fact the
bond strength of a chemical qualitatively. The \textit{vibrational}
Raman frequency can be determined by the first-order response,
because it lies on the frequency $\omega$ of the optical phonon at
the Brillouin-zone center ($\Gamma$ point).\cite{R13,R14,R15} Raman
intensity of the Raman mode with the Raman frequency $\omega$ can be
calculated under the Placzek approximation.\cite{R14} Raman spectra
can be calculated by the \textsc{PWSCF} implementation within the
density functional perturbation theory (DFPT) formalism.\cite{R14}

In the previous works, twice of the band gap or band
energy,\cite{R06} the reference energy,\cite{R07} and
electron-holding energy\cite{R08} were used to define the resistant
force of bond. We introduce the Raman frequency of a
\textit{vibrational} Raman active bond, $\omega _m$, to embody the
resistant force $F_m$ of this bond as

\noindent
\begin{equation}
\label{eq1} F_m \varpropto \omega _m \exp (\omega _m / \omega _0),
\end{equation}

\noindent where $\omega_0$ is a constant. The suffix $m$ indicates
the ordinal number of the \textit{vibrational} Raman mode. Inasmuch
as hardness of a material is attributed to the collective
contribution of all the chemical bonds in any direction, hardness
can be described by the \textit{compromise} resistant force. Due to
differences in the resistant force among different types of bonds,
the \textit{compromise} resistant force provided by all the chemical
bonds with the \textit{vibrational} Raman active property should be
depicted by a \textit{weighted geometric} average \cite{R06} of all
the \textit{vibrational} Raman frequencies. Thus the contributions
of all the chemical bonds can be equivalent to that of a single
\textit{isotropic} chemical bond with an imaginary Raman frequency
$\omega_{d}$, here we call $\omega_{d}$ the \textit{diagnostic Raman
frequency}, which is expressed as

\noindent
\begin{equation}
\label{eq2} \omega _d = \left[ {\prod {\left( {\omega _m }
\right)^{I_m }} } \right]^{1 \mathord{\left/ {\vphantom {1 {\sum
{I_m } }}} \right. \kern-\nulldelimiterspace} {\sum {I_m } }}.
\end{equation}

\noindent Here $I_{m}$ is the relative intensity of the $m$th
\textit{vibrational} Raman mode with the Raman frequency $\omega_m$
and indicates its contribution fraction or the weighted factor.
Therefore, the single equivalent \textit{isotropic} chemical bond
provides the resistant force $F$ should be

\noindent
\begin{equation}
\label{eq3} F \varpropto \omega _d \exp (\omega _d / \omega _0).
\end{equation}

\noindent It is clear that $F \to 0$ when $\omega_{d} \to 0$.
Because hardness is understood as the resistance offered by a given
material to applied mechanical action, the Vickers hardness can be
expressed as $H_{V}$ = $A F$, where $A$ is a proportional constant.
Despite the fact that we cannot give the ab initio deduction for the
exact expressions of $A$ and $\omega_{0}$ from the theory, we can
still obtain their values based on the semi-empirical method. It is
well known that the zincblende crystals have the simplest Raman
spectra with two Raman modes (one LO and one TO) and the zincblende
diamond is the hardest material so far. Therefore, we can use the
known Raman spectra and hardness for the typical zinc-blende
covalent crystals, to determine the constants of $A$ and
$\omega_{0}$. From $H_{V}$ against $\omega_{d}$ for the first 14
typical zincblende covalent crystals from Table I as shown in Fig.
1, we obtain

\noindent
\begin{equation}
\label{eq4} H_V {\rm (GPa)} = 0.011 \omega _d \exp (\omega _d /
704.8),
\end{equation}

\noindent where $\omega_{d}$ is in the unit of cm$^{-1}$. For all
the 22 crystals listed in Table I, the Vickers hardness values
calculated by the above formula are given. Our results are in good
agreement with the experimental data and the theoretical values by
the other models.

To confirm the universality of Eq. (4), we first apply it to the
wurtzite crystals with two types of chemical bonds. The results
validate the good agreements between the calculated and experimental
values as shown by circles in Fig. 2.

Next we focus on some typical complex crystals as listed in Table
II. For $\beta$-Si$_{3}$N$_{4}$ (a fourfold coordinated Si atom is
linked by four threefold coordinated N atoms), from the seven
vibrational Raman modes measured in experiment,\cite{R19} we obtain
$\omega_{d} = 818.9$ cm$^{-1}$ by Eq. (2) and then $H_{V} = 28.8$
GPa by Eq. (4). The calculated Vickers hardness is in good agreement
with the experimental value of 30 GPa and the theoretical value of
30.3 GPa.\cite{R06} For the experimental Raman spectra of
$\alpha$-SiO$_{2}$ in Ref. 14 and Stishovite in Ref. 20, the
theoretical analyses reveal that the strong Raman mode at 207
cm$^{-1}$ for $\alpha$-SiO$_{2}$ and the two strong Raman modes at
589 and 231 cm$^{-1}$ for Stishovite should be excluded for
evaluating hardness, because they belong to the rotational Raman
modes. When other very weak Raman modes are ignored simultaneously,
only two relatively strong Raman modes need to be included for
calculating hardness: 464 and 450 cm$^{-1}$ for $\alpha$-SiO$_{2}$
in Ref. 14 as well as 967 and 753 cm$^{-1}$ for Stishovite in Ref.
20. The calculated values are 9.8 GPa for $\alpha$-SiO$_{2}$ and
25.3 GPa for Stishovite, respectively, which are in good agreement
with the experimental values of 11 GPa in Ref. 6 for
$\alpha$-SiO$_{2}$ and of 17-23 GPa in Ref. 5 (but lower than 33 GPa
in Refs. 5 and 6) for Stishovite. For $\alpha$-SiO$_{2}$, the
Vickers hardness of 30.6 GPa predicted by the microscopic model is
remarkably higher than the experimental value of 11 GPa in Ref. 6,
whereas our predicated value of 9.8 GPa is very close to the
experimental value.

We now turn to explore the boron-rich systems, including
$\alpha$-Boron, B$_{4}$C and B$_{6}$O. Based on both the theoretical
and experimental analyses,\cite{R21,R22,R23,R24,R25} Vast \textit{et
al}. have pointed out that the Raman modes of $\alpha$-Boron could
be classified into three groups: intericosahedral modes at high
frequency, intraicosahedral modes at middle frequency, and
librational modes at low frequency.\cite{R22} This opinion is also
feasible to B$_{4}$C and B$_{6}$O. Lazzari \textit{et al}. have
validated the two measured low-frequency Raman modes (498 and 534
cm$^{ - 1})$ for B$_{4}$C are the rotational and librational modes,
respectively. \cite{R24} For $\alpha$-Boron, from the experimental
Raman spectrum, \cite{R21} we evaluate the Vickers hardness to be
39.8 GPa ($\omega_{d} = 945.8$ cm$^{-1}$), which agrees with the
experimental value of 42 GPa. \cite{R25} By using our present model
combined with the measured Raman spectra, \cite{R21,R23} the
calculated hardness are 40.2 GPa ($\omega_{d} = 950.9$ cm$^{-1}$)
for B$_{4}$C and 44.9 GPa ($\omega _{d} = 994.6$ cm$^{-1}$) for
B$_{6}$O, which are in good agreement with the experimental values
of 42-49 GPa for B$_{4}$C in Ref. 5 and 45 GPa for B$_{6}$O in Ref.
25, respectively.

Finally, we concern the ternary superhard BC$_2$N. Recently, we
reported a new phase ($z$-BC$_2$N) with the P-42M space group,
\cite{R36} and the simulated XRD pattern is in good agreement with
the experimental data.\cite{R11} However, the first-principles
calculation reveals that the non-resonant first-order Stokes Raman
spectrum of $z$-BC$_2$N does not match to the measured Raman pattern
of the synthesized BC$_2$N sample. \cite{R12} We construct a
modified structure of BC$_2$N, named as $z^*$-BC$_2$N here, with the
same space group as $z$-BC$_2$N. In fact, the unique difference
between $z^*$-BC$_2$N and $z$-BC$_2$N is that only B and N atoms are
interchanged each other. The simulated Raman spectrum of
$z^*$-BC$_2$N is in good agreement with the experimental Raman
pattern of the synthesized BC$_2$N sample. \cite{R12} Based on the
simulated Raman spectrum of $z^*$-BC$_2$N as Table II, the evaluated
hardness of $z^*$-BC$_2$N is 75.5 GPa, which agrees with the
experimental value of 76 GPa. \cite{R11}

It should be noted that the role of $d$ valence electron in chemical
bond are not considered here. This model could also be simplified
for roughly estimating hardness of potential superhard materials
with the Vickers hardness above 40 GPa, as follows: to avoid the
procedure eliminating the translational and rotational Raman modes,
we only need to select the strong Raman modes with their Raman
frequencies higher than the critical frequency $\omega_{c} = 687$
cm$^{-1}$ ($\omega_{c}$ is the diagnostic Raman frequency of $H_{V}
= 20$ GPa, when a substance with its Vickers hardness between 20 and
40 GPa is usually considered as a hard material). For example, for
superhard diamond-like BC$_5$ ($d$-BC$_5$),\cite{R37} there is one
strongest Raman mode located at 1200 cm$^{-1}$ above $\omega_{c}$ in
the measured Raman spectrum. By using this single strongest Raman
mode, we estimate its Vickers hardness to be 72.4 GPa, which is in
excellent agreement with the experimental value of 71 GPa.\cite{R37}
To display an intuitive comparison, Fig. 2 summaries the calculated
and experimental Vickers hardness values for 25 crystals. The good
agreement validates the predictive power of our model.

In conclusion, we propose a model that reveals a quantitative
relationship between hardness and vibrational Raman frequencies.
This model has a tremendous advantage that, to evaluate the hardness
synthesized materials, we do not need to know the exact atom
arrangement provided that the Raman spectrum can be measured. The
present work validates a universal technique for the nondestructive
and indirect hardness measurement, it also be potential used to
explain enhancement of the surface or vacancy hardness because of
the high sensitivity of the Raman spectra to the small change in
structure. Moreover, since different Raman configurations for a
single crystal will result in different Raman spectra, our model can
be anticipated to explore the anisotropy of the hardness to some
extent. This Raman model finds in fact a new application for the
Raman spectroscopy.

This work is in part supported by the National Natural Science
Foundation of China under Grants Nos. 50821001 and 50532020, by
the 973 Program of China under Grant Nos. 2006CB921805 and
2005CB724400.


\newpage

\noindent Table I. Hardness, $\omega _{m}$ and $I_{m}$ (in
parenthesis) of LO and TO Raman modes, for 22 kinds of diamond-like
crystals. $H_V^{Our}$, $H_V^{Gao}$ and $H_V^{Sim}$ are the
calculated Vickers hardness by our, Gao's [6] and \v{S}imunek's [7]
models, respectively. $H_V^{Exp}$ is the experimental Vickers
hardness (unless noted, from Refs. [6,7]). Despite the fact that
there are no measured Vickers hardness for 8 materials from ZnSe and
below, their Knoop hardness values marked by the asterisk are listed
as a reference.

\begin{table}[htbp]
\begin{tabular}{lcccccc}
\hline\hline Crystals & $\omega_{\rm LO}$ & $\omega _{\rm TO}$ &
$H_V^{Our}$ & $H_V^{Gao}$ & $H_V^{Sim}$ &
$H_V^{Exp}$ \\
\hline
Diamond& 1332 (1.0)$^{\rm a}$& 1332 (1.0)$^{\rm a}$& 97.0 &
93.6 & 95.4 & 96 \\
BN & 1305 (0.7)$^{\rm b}$ & 1055 (1.0)$^{\rm b}$ & 64.9 & 64.5 &
63.2 & 66 \\
SiC & 976 (0.6)$^{\rm c}$ & 796 (1.0)$^{\rm c}$ & 32.0 & 30.3 & 31.1
& 34 \\
BP & 829 (1.0)$^{\rm b}$ & 799 (0.2)$^{\rm b}$ & 29.2 & 31.2 & 26.0
& 33 \\
AlN & 893 (0.2)$^{\rm b}$ & 668 (1.0)$^{\rm b}$ & 20.9 & 21.7 & 17.6
& 18 \\
GaN & 741 (1.0)$^{\rm d}$ & 555 (0.6)$^{\rm d}$ & 18.8 & 18.1 & 18.5
& 15.1 \\
Si & 520 (1.0)$^{\rm a}$ & 520 (1.0)$^{\rm a}$ & 12.0 & 13.6 & 11.3
& 12 \\
AlP & 501$^{\rm e}$ (0.5) & 440$^{\rm e}$ (1.0) & 9.7 & 9.6 & 7.9 &
9.4 \\
InN & 588 (1.0)$^{\rm d}$ & 457 (0.6)$^{\rm d}$ & 12.6 & 10.4 & 8.2
& 9 \\
Ge & 304 (1.0)$^{\rm a}$ & 304 (1.0)$^{\rm a}$ & 5.2 & 11.7 & 9.7 &
8.8 \\
GaAs & 292 (0.6)$^{\rm f}$ & 269 (1.0)$^{\rm f}$ & 4.5 & 8.0 & 7.4 &
6.8$^{\rm m}$ \\
InP & 345 (0.9)$^{\rm f}$ & 304 (1.0)$^{\rm f}$ & 5.6 & 6.0 & 5.1 &
5.4 \\
InAs & 238$^{\rm g}$ (0.5) & 217$^{\rm g}$ (1.0) & 3.4 & 5.7 & 4.5
& 3.8 \\
ZnSe & 252 (1.0)$^{\rm h}$ & 203 (0.7)$^{\rm h}$ & 3.5 & - & 2.6 &
1.1$^{\rm m}$ \\
BAs & 714 (0.4)$^{\rm i}$ & 695 (1.0)$^{\rm i}$ & 20.8 & - & 19.9 &
19* \\
GaP & 403 (1.0)$^{\rm e}$ & 367 (0.2)$^{\rm e}$ & 7.7 & 8.9 & 8.7 &
9.5* \\
AlAs & 404$^{\rm j}$ (0.5) & 363$^{\rm j}$ (1.0) & 7.2 & 8.5 & 6.8 &
5* \\
GaSb & 237 (0.7)$^{\rm a}$ & 227 (1.0)$^{\rm a}$ & 3.5 & 6.0 & 5.6 &
4.5* \\
AlSb & 340 (0.3)$^{\rm f}$ & 319 (1.0)$^{\rm f}$ & 5.6 & 4.9 & 4.9
& 4* \\
InSb & 189 (0.7)$^{\rm a}$ & 179 (1.0)$^{\rm a}$ & 2.6 & 4.3 & 3.6 &
2.2* \\
ZnS & 351 (1.0)$^{\rm k}$ & 276 (0.1)$^{\rm k}$ & 6.2 & - & 2.7 &
1.8* \\
ZnTe & 206 (1.0)$^{\rm l}$ & 177 (0.8)$^{\rm l}$ & 2.8 & - & 2.3 &
1.0* \\
\hline\hline
\end{tabular}
\end{table}

\noindent $^{\rm a}$Reference [16]. $^{\rm b}$Reference [17]. $^{\rm
c}$Reference [18]. $^{\rm d}$Reference [26]. $^{\rm e}$Reference
[27]. $^{\rm f}$Reference [28]. $^{\rm g}$Reference [29]. $^{\rm
h}$Reference [30]. $^{\rm i}$Reference [31]. $^{\rm j}$Reference
[32]. $^{\rm k}$Reference [33]. $^{\rm l}$Reference [34]. $^{\rm
m}$Reference [35].

\newpage
\noindent Table II. For 7 kinds of complex materials,
$\omega_{m}$($I_{m}$) of vibrational Raman modes, our Vickers
hardness $H_V^{Our}$, and the experimental Vickers hardness
$H_V^{Exp}$. The Vickers hardness $H_V^{Gao}$ predicated by the
microscopic model has been also given as a comparison.

\begin{table}[htbp]
\centering
\begin{tabular}{lllllllllll}
\hline\hline
Crystals & $\omega_{1}$($I_{1}$) & $\omega_{2}$($I_{2}$)
& $\omega_{3}$($I_{3}$) & $\omega_{4}$($I_{4}$) &
$\omega_{5}$($I_{5}$) &
$\omega_{6}$($I_{6}$) & $\omega_{7}$($I_{7}$) & $H_V^{Our}$ & $H_V^{Gao}$ & $H_V^{Exp} $ \\
\hline
$\beta$-Si$_{3}$N$_{4}$ & 1047(1.0)$^{\rm a}$ & 939(1.0)$^{\rm
a}$ & 928(0.8)$^{\rm a}$ & 865(0.6)$^{\rm a}$ & 732(1.0)$^{\rm a}$ &
619(0.4)$^{\rm a}$
& 451(0.6)$^{\rm a}$ & 28.8 & 30.3$^{\rm f}$ & 30$^{\rm f}$ \\
$\alpha$-SiO$_{2}$ & 464(1.0)$^{\rm b}$ & 450(0.1)$^{\rm b}$ & &
& & & & 9.8 & 30.6$^{\rm f}$ & 11$^{\rm f}$ \\
Stishovite & 967(0.1)$^{\rm c}$ & 753(1.0)$^{\rm c}$ & & & & & &
25.3 & 30.4$^{\rm f}$ & 17-23$^{\rm g}$, 33$^{\rm g}$ \\
$\alpha$-Boron & 1186(0.85)$^{\rm d}$ & 1123(0.12)$^{\rm d}$ &
933(0.7)$^{\rm d}$ & 795(1.0)$^{\rm d}$ & 776(0.15)$^{\rm d}$ & & &
39.8 & 42$^{\rm h}$& 42$^{\rm h}$ \\
B$_{4}$C & 1085(1.0)$^{\rm d}$ & 1000(0.5)$^{\rm d}$ &
830(0.3)$^{\rm d}$ & 720(0.4)$^{\rm d}$ & & & & 40.5 & 42$^{\rm
h}$&30$^{\rm g}$, 42-49$^{\rm g}$ \\
B$_{6}$O & 1119(0.5)$^{\rm e}$ & 1034(1.0)$^{\rm e}$ &
902(1.0)$^{\rm e}$ & & & & & 44.9 & 44$^{\rm h}$ & 38$^{\rm h}$,45$^{\rm h}$ \\
$z^*$-BC$_2$N & 1328(0.9) & 1326(1.0) & 1292(0.3) & 1176(0.6) &
1076(0.4) & 930(0.4) & & 75.5 & 78$^{\rm f}$& 76$^{\rm f}$ \\
\hline\hline
\end{tabular}
\end{table}

\noindent $^{\rm a}$Reference [19]. $^{\rm b}$Reference [14]. $^{\rm
c}$Reference [20]. $^{\rm d}$Reference [21]. $^{\rm e}$Reference
[23]. $^{\rm f}$Reference [6]. $^{\rm g}$Reference [5]. $^{\rm
h}$Reference [25].

\newpage

\noindent \textbf{Figure Captions}

\begin{figure}[h!]
\includegraphics[width=0.9\textwidth]{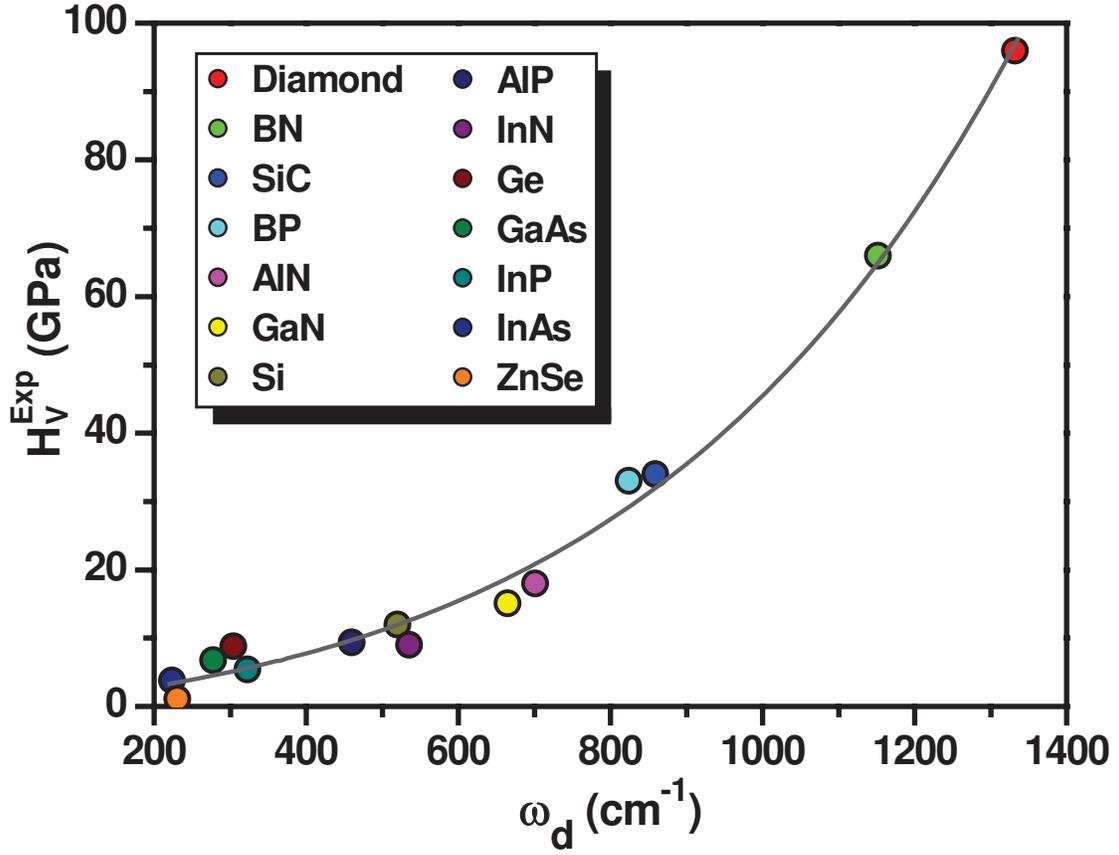}
\caption{\label{Fig} (Color online) Hardness of diamond-like crystals as
a function of diagnostic Raman frequency. The solid line is from Eq.
(4).}
\end{figure}
\newpage
\begin{figure}[h!]
\includegraphics[width=0.9\textwidth]{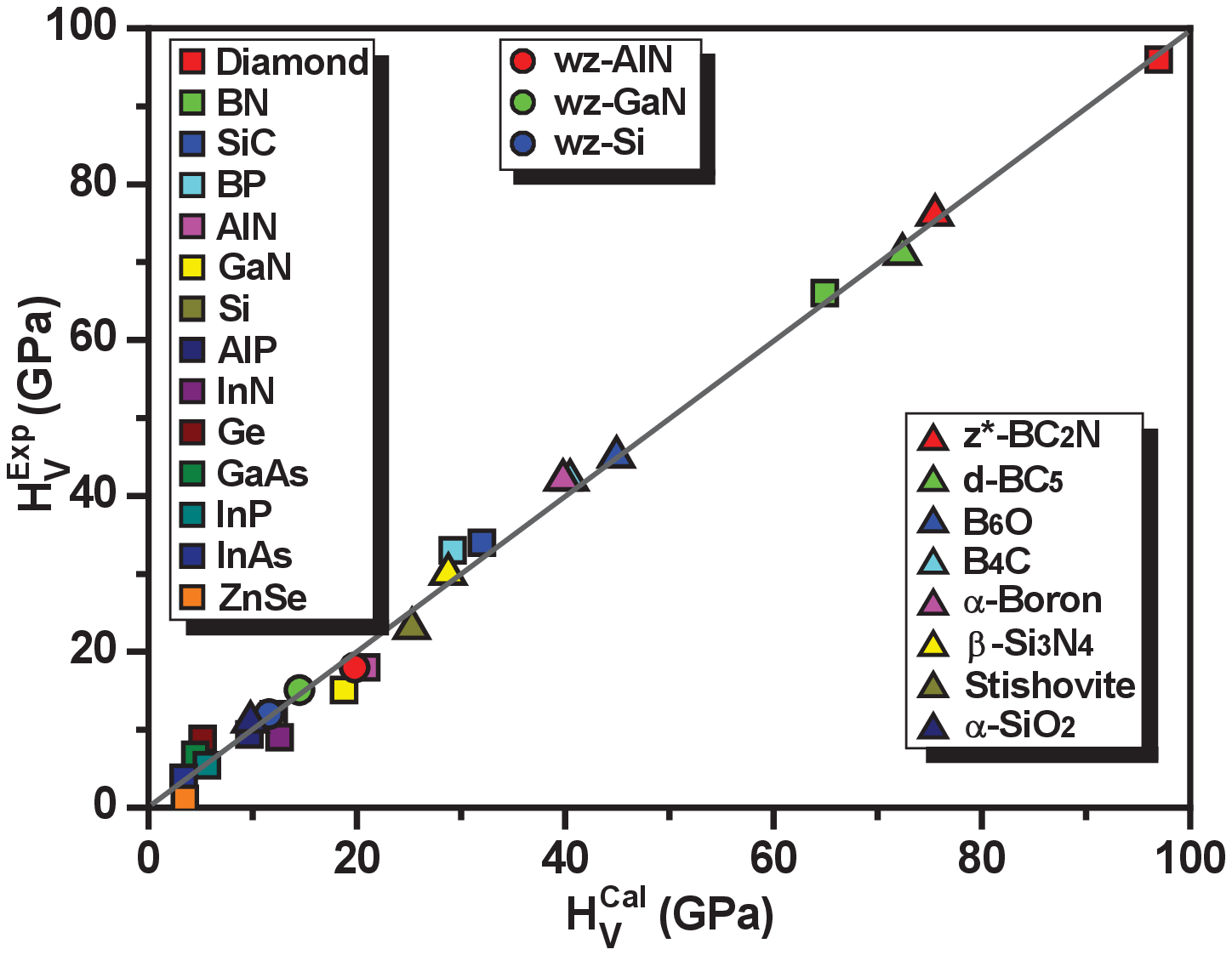}
\caption{\label{Fig} (Color online) Comparison of calculated and
experimental Vickers hardness for 25 crystals. Filled squares,
circles, and triangles denote the zinc blende, wurtzite (wz), and
complex crystals, respectively.}
\end{figure}
\newpage



\begin{references}

\bibitem{R01} R. B. Kaner, J. J. Gilman, and S. H. Tolbert, Science
\textbf{308}, 1268 (2005).

\bibitem{R02} J. Haines, J. M. L\'{e}ger, and G. Bocquillon, Annu. Rev. Mater.
Res. \textbf{31}, 1 (2001).

\bibitem{R03} A. L. Liu and M. L. Cohen, Science \textbf{245}, 841 (1989).

\bibitem{R04} D. M. Teter, Mater. Res. Bull. \textbf{23}, 22 (1998).

\bibitem{R05} V. V. Brazhkin, A. G. Lyapin, and R. J. Hemley, Philos. Mag. A
\textbf{82}, 231 (2002).

\bibitem{R06} F. M. Gao, J. L. He, E. D. Wu, S. M. Liu, D. L. Yu, D. C. Li, S.Y.
Zhang, and Y. J. Tian, Phys. Rev. Lett. \textbf{91}, 015502 (2003)
and references therein.

\bibitem{R07} A. \v{S}im$\mathop {\rm u}\limits^\circ$nek and J. Vack\'{a}\v{r},
Phys. Rev. Lett. \textbf{96}, 085501 (2006) and references therein.

\bibitem{R08} K. Y. Li, X. T. Wang, F. F. Zhang, and D. F. Xue, Phys. Rev. Lett.
\textbf{100}, 235504 (2008).

\bibitem{R09} A. Zerr, G. Miehe, G. Serghiou, M. Schwarz, E. Kroke, R. Riedel, H.
Fue{\ss}, P. Kroll, and R. Boehler, Nature \textbf{400}, 340 (1999).

\bibitem{R10} A. F. Young, C. Sanloup, E. Gregoryanz, S. Scandolo, R. J. Hemley, and H. K.
Mao, Phys. Rev. Lett. \textbf{96}, 155501 (2006).

\bibitem{R11} V. L. Solozhenko, D. Andrault, G. Fiquet, M. Mezouar, and D. Rubie,
Appl. Phys. Lett. \textbf{78}, 1385 (2001).

\bibitem{R12} H. W. Hubble, I. Kudryashov, V. L. Solozhenko, P. V. Zinin, S. K.
Sharma, and L. C. Ming, J. Raman Spectrosc. \textbf{35}, 822 (2004).

\bibitem{R13} S. Baroni, S. de Gironcoli, A. D. Corso, and P. Giannozzi, Rev. Mod. Phys.
\textbf{73}, 515 (2001); X. Gonze and C. Lee, Phys. Rev. B
\textbf{55}, 10355 (1997).

\bibitem{R14} M. Lazzeri and F. Mauri, Phys. Rev. Lett. \textbf{90}, 036401
(2003); www.pwscf.org.

\bibitem{R15} S. Baroni and R. Resta, Phys. Rev. B \textbf{33}, 5969
(1986); P. Umari, X. Gonze, and A. Pasquarello, Phys. Rev. Lett.
\textbf{90}, 027401 (2003).

\bibitem{R16} A. Kailer, K. G. Nickel, and Y. G. Gogotsi, J. Raman Spectrosc. \textbf{30},
939 (1999).

\bibitem{R17} J. A. Sanjurjo, E. L\'{o}pez-Cruz, P. Vogl, and M. Cardona, Phys. Rev. B
\textbf{28}, 4579 (1983).

\bibitem{R18} Z. C. Feng, A. J. Mascarenhas, W. J. Choyke, and J. A. Powell, J. Appl.
Phys. \textbf{64}, 3176 (1988).

\bibitem{R19} N. Wada, S. A. Solin, J. Wong, and S. Prochazka, J. Non-Cryst. Solids
\textbf{43}, 7 (1981).

\bibitem{R20} R. J. Hemley, H. K. Mao, and E. C. T. Chao, Phys. Chem. Minerals.
\textbf{13}, 285 (1986).

\bibitem{R21} D. R. Tallant, T. L. Aselage, A. N. Campbell, and D. Emin, Phys. Rev. B
\textbf{40}, 5649 (1989).

\bibitem{R22} N. Vast, S. Baroni, G. Zerah, J. M. Besson, A. Polian, M.
Grimsditch, and J. C. Chervin, Phys. Rev. Lett. \textbf{78}, 693
(1997).

\bibitem{R23} H. Werheit, and U. Kuhlmann, J. Solid. State. Chem. \textbf{133},
260 (1997).

\bibitem{R24} R. Lazzari, N. Vast, J. M. Besson, S. Baroni, and A. D. Corso, Phys.
Rev. Lett. \textbf{83}, 3230 (1999).

\bibitem{R25} J. L. He, E. D. Wu, H. T. Wang, R. P. Liu, and Y. J. Tian, Phys.
Rev. Lett. \textbf{94}, 015504 (2005) and references therein.

\bibitem{R26} A. Tabata, A. P. Lima, L. K. Teles, L. M. R. Scolfaro, J. R. Leite,
V. Lemos, B. Schttker, T. Frey, D. Schikora, and K. Lischka, Appl.
Phys. Lett. \textbf{74}, 362 (1999).

\bibitem{R27} G. D. Mahan, R. Gupta, Q. Xiong, C. K. Adu, and P. C. Eklund, Phys.
Rev. B \textbf{68}, 073402 (2003).

\bibitem{R28} A. Mooradian and G. B. Wright, Solid State Commun. \textbf{4}, 431
(1966).

\bibitem{R29} R. Carles, N. Saint-Cricq, J. B. Renucci, M. A. Renucci, and A.
Zwick, Phys. Rev. B \textbf{22}, 4804 (1980).

\bibitem{R30} B. Y. Geng, Q. B. Du, X. W. Liu, J. Z. Ma, X. W. Wei, and L. D.
Zhang, Appl. Phys. Lett. \textbf{89}, 033115 (2006).

\bibitem{R31} R. G. Greene, H. Luo, A. L. Ruoff, S. S. Trail, and F. J. DiSalvo,
Jr., Phys. Rev. Lett. \textbf{73}, 2476 (1994).

\bibitem{R32} J. Wagner, A. Fischer, W. Braun, and K. Ploog, Phys. Rev. Lett.
\textbf{49}, 7295 (1994).

\bibitem{R33} O. Brafman and S. S. Mitra, Phys. Rev. \textbf{171}, 931 (1968).

\bibitem{R34} R. Vogelgesang, A. J. Mayur, M. Dean Sciacca, E. Oh, I. Miotkowski,
A. K. Ramdas, S. Rofriguez, and G. Bauer, J. Raman Spectrosc.
\textbf{27}, 239 (1996).

\bibitem{R35} I. Yonenaga, Chemistry for Sustainable Development \textbf{9}, 19
(2001).

\bibitem{R36}X. F. Zhou, J. Sun, Y. X. Fan, J. Chen, H. T. Wang, X. J. Guo,
J. L. He, and Y. J. Tian, Phys. Rev. B \textbf{76}, 100101(R)
(2007); X. F. Zhou, J. Sun, G. R. Qian, X. J. Guo, Z. Y. Liu, Y. J.
Tian, and H. T. Wang, J. Appl. Phys. \textbf{105}, 093521 (2009).

\bibitem{R37} V. L. Solozhenko, O. O. Kurakevych, D. Andrault, Y. L. Godec, and M. Mezouar,
Phys. Rev. Lett. \textbf{102}, 015506 (2009).

\end{references}
\end{document}